\documentclass[aps,prb,showpacs,amsmath,twocolumn,amssymb,superscriptaddress,letterpaper]{revtex4}
\usepackage{mathrsfs}
\usepackage{graphicx}
\usepackage{graphics}
\usepackage{subfigure}
\usepackage{bm}
\usepackage{dcolumn}
\usepackage{amsmath,bm}
\usepackage{color}

\begin{document}
\title{Surface plasmon polaritons in topological insulator}
\author{Junjie Qi}
\affiliation{ Institute of Physics, Chinese Academy of Sciences,
Beijing 100190, China}
\affiliation{ Collaborative Innovation Center of Quantum Matter, Beijing, China}
\author{Haiwen Liu}
\email{haiwen.liu@pku.edu.cn}
\affiliation{ Collaborative Innovation Center of Quantum Matter, Beijing, China}
\affiliation{International Center for Quantum Materials and School of Physics, Peking
University, Beijing 100871, China}

\author{X. C. Xie}
\affiliation{ Collaborative Innovation Center of Quantum Matter, Beijing, China}
\affiliation{International Center for Quantum Materials and School of Physics, Peking
University, Beijing 100871, China}

\date{\today}

\begin{abstract}
We study surface plasmon polaritons on topological insulator-vacuum interface. When the time-reversal symmetry is broken due to ferromagnetic coupling,  the surface states exhibit magneto-optical Kerr effect. This effect gives rise to a novel transverse type surface plasmon polariton, besides the longitudinal type. In specific, these two types contain three different channels, corresponding to the pole of determinant of Fresnel reflection matrix. All three channels of surface plasmon polaritons display tight confinement, long lifetime and show strong light-matter coupling with a dipole emitter.
\end{abstract}

\pacs{73.20.Mf, 71.45.Gm, 85.75.-d}

\maketitle

\section{Introduction}

Surface plasmon polaritons (SPPs) describe electromagnetic collective modes of electrons propagating at the interface between a conductor and a dielectric, which are evanescently confined in the perpendicular direction. SPPs show potential applications in the fields of subwavelength optics,\cite{Bozhevolnyi2001,Maier2003,Bozhevolnyi2006, Nagpal2009,Lezec2002,Fang2011} near-field optics,\cite{Melville2004,Kim2008,Smythe2009} and photonic data storage.\cite{Barnes2003} However, even noble metals with high conductivities (e.g. gold or silver), which are widely regarded as traditional plasmonic materials, suffer from large Ohmic losses at high frequencies.\cite{Johnson,Marton} The large losses hinder the development of devices based on SPPs, thus materials with low losses are required.\cite{Ebbesen2003}
Recently, researchers have shown that Dirac materials (e. g. graphene) are suitable candidates for plasmonic applications, with lower loss compared to noble metals.\cite{Neto2009,Feng2011} Besides, the SPPs in Dirac materials have a number of advantages such as frequency tunability with gate voltage, tight confinement due to shorter wavelength, long lifetime, strong light-matter coupling.\cite{Koppens2011, Novoselov2012,Stauber2013} Recently, experimentalists have  successfully detected the real-space images of plasmon in tailored graphene nanostructures by
infrared nano-imaging techniques.\cite{Fei2012,ChenJ2012} Recent experimental verification of tunability of graphene plasmons have greatly stimulated the research interests of plasmons in Dirac systems.\cite{Chaoxing2013,Stauber2013b}

\begin{figure}[htbp]
  \centering
  \includegraphics[width=9.0cm]{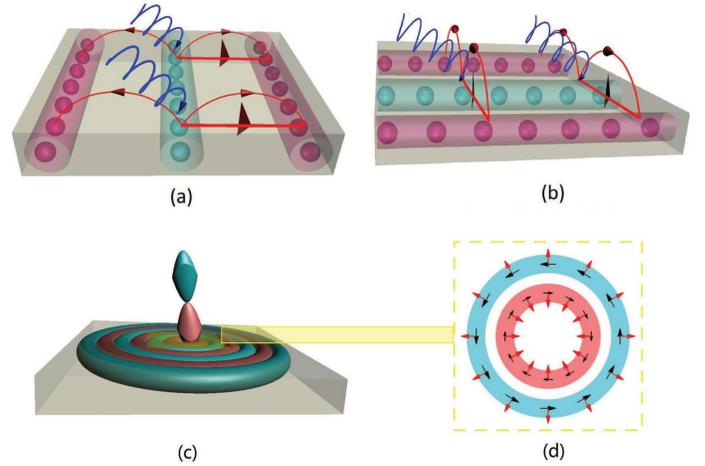}\\
  \caption{(Color online) Schematic plot of SPPs and SPP-emitter coupling on gapped TI surface. (a)The longitudinal SPP propagates parallel to the incident plane of light. (b)Due to the magneto-optical Kerr effect, the transverse SPP propagates perpendicular to the incident plane of light. The blue curved arrows denote the incident light, and the red arrows denote the propagation of SPPs, and the blue and red sphere represent different charges.  (c) Near electric field produced by a dipole emitter (placed 10nm away from surface) stimulate SPPs on gapped TI  surface. (d) The schematic propagation direction of SPP in the SPP-emitter setup. The radial SPP is marked by red arrows, which is similar to SPP in traditional plasmonic materials. The tangential SPP is caused by the magneto-optical Kerr effect, marked by the black arrows.}
\end{figure}

Topological insulators(TIs) are a new class of Dirac systems with both insulating bulk and metallic surface (edge) states which attract great interest in recent years.\cite{Bernevig2006,Fu2007,Konig2007,Qi2008,Zhang2009}  The angle-resolved photoemission spectroscopy(ARPES) experiments have demonstrated that Bi$_{x}$Sb$_{1-x}$, Bi$_{2}$Se$_{3}$ and Bi$_{2}$Te$_{3}$ are three dimentional TIs. \cite{Heish2008,Heish2009,Hsieh2010,Noh2008,Xia2009,Hsieh2009,Chen2009} The gapless surface states in 3D TIs are topological protected by the time-reversal symmetry(TRS).\cite{Kane2010}These gapless TI surface states possesses many peculiar characteristics, such as spin-momentum locking mechanism and topological magneto-electric effect.~\cite{ZhangSC2010} Specifically, the spin-momentum locking mechanism of helical surface states leads to coupling transport of spin and charge,\cite{Burkov} and novel spin-plasmon besides the charge-plasmon.\cite{Raghu2010,Chaoxing2013} Moreover, a gap-opening term can be acquired by breaking the TRS\cite{Chen2010} and leads to strong magneto-optical effect in these systems.\cite{Tse2010}

In this paper, we investigate the coupling features between a dipole emitter and gapped TIs surface states, and study the properties of SPPs which are bound on a gapped TI-vacuum interface. There are two specific characteristics of SPPs in the TRS breaking TIs surface states. Firstly, we find a novel transverse type of SPPs besides the traditional longitudinal type. In specific, the s-polarization light can lead to a transverse SPP due to the magneto-optical Kerr effect, as shown in Fig.1 (b). Besides, the p-polarization light can excite a longitudinal SPP, similar to SPP in traditional plasmonic meterials, as shown in Fig.1 (a). These two types of SPPs exhibit very short confinement (1-5 nm), relatively long propagation length. Secondly, we consider the coupling between a single dipole emitter and the SPPs modes, as shown in Fig.1 (c)-(d). The decay rate can be enhanced by $\sim10^7$ times compared to free space, due to the short confinement of SPPs. The high strength of coupling with the light makes TI materials be good platform for strong light-matter interactions.

The structure of our paper is organized as follows. In Sec. II, we derive the expression of the optical conductivity from the massive Dirac Hamiltonian. In Sec. III, we consider the magneto-optical Kerr effect when applying an electric field to TI surface. Further, we work out the Fresnel reflected matrix. In Sec IV, we obtain the dispersions of SPPs. In Sec. V, we study the light-SPP coupling in this system. Finally, we end our analysis with a conclusion in Sec. VI.

\section{Model}

 We consider the top TI surface states with breaking TRS, which is described by a massive Dirac Hamiltonian \cite{Kane2010}

\begin{equation}
H\left(\mathbf{k}\right)=\hbar v_{f}\left(\mathbf{\hat{z}}\times\mathbf{k}\right)\cdot\mathbf{\tau}+\Delta\tau_{z}
\end{equation}
where $\mathbf{k}=(k_{x},k_{y})$ is the wave vector  and $\textbf{$\tau$}=(\tau_{x},\tau_{y},\tau_{z})$ are Pauli matrices describing electron spin. The first term in Hamiltonian describes the gapless linear dispersion of the TI surface states. To break TRS, we introduce a mass term $\Delta \tau_{z}$ as the second term in Eq.(1). The mass term can be generated by ferromagnetic coupling with an insulating magnetic material.\cite{Chen2010} In the following part, we only focus on the SPPs on the top TI surface.

The optical properties of the surface state is determined by optical conductivity. For massive Dirac Hamiltonian, we consider both the longitudinal conductivity $\sigma_{xx}$ and  Hall conductivity $\sigma_{xy}$ in this system. We assume that the mass term $\triangle>0$ and set the Fermi level $E_{F}$ above the gap $E_{F}>\triangle$ . Utilizing $e^{2}/\hbar^{2}$ as unit in the following equations, the optical conductivities are expressed by: \cite{Tse2010}

\begin{equation}
\begin{aligned}
\sigma_{xx}(\omega)=&\frac{E_{F}}{4\pi}\frac{i}{\omega+i\tau^{-1}}+(\frac{1}{16}+(\frac{\triangle}{2\omega})^{2})\{\theta(\omega-2E_{F})
\\&+\frac{i}{\pi}log|\frac{\omega-2E_{F}}{\omega+2E_{F}}|\}
\end{aligned}
\end{equation}
\begin{figure}[htbp]
  \centering
  \includegraphics[width=8.0cm]{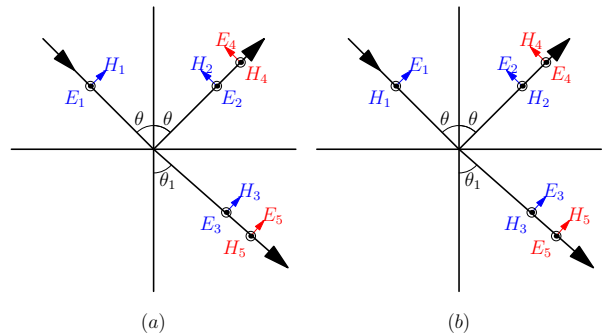}\\
  \caption{(Color online) Diagram of reflection and refraction at the interface between the TI surface and vacuum. (a)The s-polarized incident wave case. (b)The p-polarized incident wave case.}
\end{figure}

\begin{equation}\small
\sigma_{xy}(\omega)=\frac{\triangle}{4}\frac{i}{\omega+i\tau^{-1}}\theta(\omega-2E_{F})+\frac{\triangle}{4\pi}\frac{1}{\omega+i\tau^{-1}}log|\frac{\omega+2E_{F}}{\omega-2E_{F}}|
\end{equation}

The first term in $\sigma_{xx}$ describes the Drude weight for the intraband transitions. Here we include a finite relaxation time $\tau$ which is experimentally determined by the surface states carrier mobility $\mu \approx 1000$ $\mathrm{cm^{2}/(V s)}$. \cite{Butch2010}  Then we get $\tau=\mu E_{F}/ev_{F}^{2}\approx10^{-12}$ s for $E_{F}=0.1$ eV , where $v_{F}\approx 10^{5}$ m/s is the Fermi velocity. We use $\tau\approx10^{-12}$ s in the following calculations. The second term is caused by interband transitions, when $\omega \geq 2E_{F}$. Hereafter, we focus on the low frequency case ($\omega<2E_{F}$), and Eq. (3) can be simplified into $\sigma_{xy}=\triangle/4\pi E_{F}$. In homogenous TIs, we have $\sigma_{xx}=\sigma_{yy}$ and $\sigma_{xy}=-\sigma_{yx}$.

\section{magneto-optical Kerr effect}

 When applying an electromagnetic wave on the gapped TI surface, the Hall conductivity $\sigma_{xy}$ induces the magneto-optical Kerr effect. To be specific, the reflected waves from the surface of TI have different polarization from the incident waves. We define the Fresnel reflection matrix $\Re$ as 

\begin{equation}
\Re=
   \left(
  \begin{array}{cc}
  r_{pp} & r_{ps} \\
  r_{sp} & r_{ss} \\
  \end{array}\right)
\end{equation}
where $s$ and $p$ denote the s-polarization and p-polarization respectively, and $r_{ij}$ ($i$,$j$=$s$,$p$) is the ratio of electric fields between the incident j-polarization wave and the reflected i-polarization wave.  We consider the low frequency case ($\omega<2E_{F}$), and $\sigma_{xy}$ is a small quantity compared to $\sigma_{xx}$. The diagonal elements $r_{pp}$ and $r_{ss}$  are:\cite{Koppens2011}

\begin{equation}
\begin{split}
  r_{ss}&\simeq \frac{ k_{\bot}-k'_{\bot}+ 4\pi\sigma_{xx} k_{0}/c}{k_{\bot}+k'_{\bot}+ 4\pi\sigma_{xx} k_{0}/c}\\
 r_{pp}& \simeq \frac{\epsilon k_{\bot}-k'_{\bot}+ 4\pi\sigma_{xx} k_{\bot}k'_{\bot}/\omega}{\epsilon k_{\bot}+k'_{\bot}+ 4\pi\sigma_{xx} k_{\bot}k'_{\bot}/\omega}
 \end{split}
\end{equation}
The off-diagonal elements $r_{sp}$ and $r_{ps}$ can be deduced by applying the standard boundary conditions. First we consider the incident s-polarization wave(Fig. 2(a)), the boundary conditions are:
\begin{equation}
\begin{split}
& E_{1} + E_{2} = E_{3} \\
& E_{4}\cos\theta = E_{5}\cos\theta_{1} \\
& H_{1}\sin\theta + H_{2}\sin\theta =  H_{3}\sin\theta_{1}\\
& E_{4}\sin\theta - \epsilon E_{5}\sin\theta_{1}= \frac{4\pi}{\omega}(k_{4}\sigma_{xx}E_{4}\cos\theta +k_{3}\sigma_{xy}E_{3})\\
& H_{1}\cos\theta - H_{2}\cos\theta = H_{3}\cos\theta_{1} + \frac{4\pi}{c}(\sigma_{yx}E_{4}\cos\theta +\sigma_{yy}E_{3})\\
& H_{4} - H_{5} = \frac{4\pi}{c}(\sigma_{xx}E_{4}\cos\theta +\sigma_{xy}E_{3})
\end{split}
\end{equation}
By using Eq. (5), we get the expression of $r_{sp}$:
\begin{equation}
  r_{sp} \simeq -\frac{(8\pi/c) \sigma_{xy}k_{\bot}k'_{\bot}}
{(k_{\bot}+k'_{\bot}+4\pi\sigma_{xx} k_{0}/c)(k'_{\bot}-\epsilon k_{\bot}- 4\pi\sigma_{xx} k_{\bot}k'_{\bot}/\omega)}
\end{equation}

Then we consider the p-polarization incident wave(Fig. 2(b)). The boundary conditions are:
\begin{equation}\small
\begin{split}
& E_{1}\cos\theta - E_{2}\cos\theta = E_{3}\cos\theta_{1} \\
& E_{4} = E_{5} \\
& H_{4}\sin\theta = H_{5}\sin\theta_{1} \\
& E_{1}\sin\theta + E_{2}\sin\theta - \epsilon E_{3}\sin\theta_{1} = \frac{4\pi}{\omega}(k_{3}\sigma_{xx}E_{3}\cos\theta_{1} +k_{5}\sigma_{xy}E_{5})\\
& H_{1} + H_{2} - H_{3} = \frac{4\pi}{c}(\sigma_{xx}E_{3}\cos\theta_{1} +\sigma_{xy}E_{5})\\
& H_{4}\cos\theta - H_{5}\cos\theta_{1} = \frac{4\pi}{c}(\sigma_{yx}E_{3}\cos\theta_{1} +\sigma_{yy}E_{5})
\end{split}
\end{equation}
thus,$r_{ps}$ is solved as
\begin{equation}
r_{ps}\simeq \frac{(8\pi/c) \sigma_{yx}k_{\bot}k'_{\bot}}
{(k_{\bot}-k'_{\bot}- 4\pi\sigma_{xx} k_{0}/c)(\epsilon k_{\bot}+k'_{\bot}+ 4\pi\sigma_{xx} k_{\bot}k'_{\bot}/\omega)}
\end{equation}
where $k_{0}=\omega/c$ is the wave vector of free space light, $k_{\parallel}$ is the in-plane wave vector, $k_{\bot}=\sqrt{k_{0}^{2}-k_{\parallel}^{2}}$, $k_{\bot}'=\sqrt{\epsilon k_{0}^{2}-k_{\parallel}^{2}}$ are the perpendicular wave vectors outside and inside the TIs respectively, and $\epsilon=100$ is the permittivity of the bulk $Bi_{2}Se_{3}$.\cite{ZhangSC2010}

\begin{figure}[htbp]
  \centering
  \includegraphics[width=8.5cm]{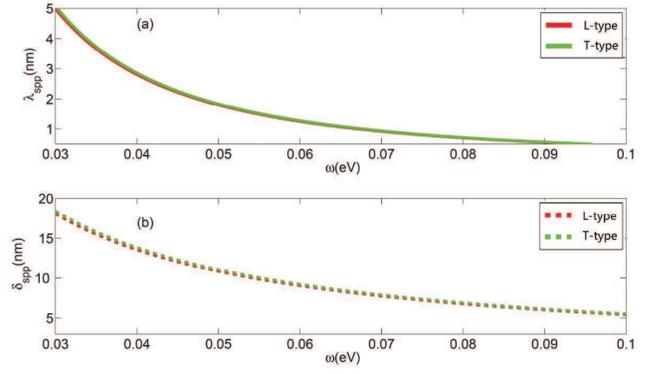}\\
  \caption{(color online) Characteristic length scales of SPPs.(a) Wave length $\lambda_{spp}$ of SPPs.(b)Propagation length $\delta_{spp}$ of SPPs. The red and green lines indicate the longitudinal type and transverse type SPPs respectively. We set Fermi energy $E_{F}$=0.05eV.}
\end{figure}

\section{Dispersion of SPPs in Gapped TIs Surface}

The pole of determinant of Fresnel reflected coefficients gives the dispersion of SPPs. Here we focus on the moderate frequency regime ($\hbar\omega\lesssim E_{F}$) and derive the dispersion of SPPs for gapped TI surface states. Normally, the s-polarization light can not produce SPPs. Previous studies have shown that s-polarization light can lead to SPP when retardation is included. However, this effect play a role in the low frequency regime ($\hbar\omega\leq\alpha E_{F}$),\cite{Stauber} where $\alpha$ is the fine structure constant. This low frequency is beyond the parameter range of our calculation, thus we neglect the effects of retardation here. We find two types of SPPs: the p-polarization incident light leads to the longitudinal type, which is corresponding to poles of $r_{pp}$ (channel-I) and $r_{ps}$ (channel-II); the s-polarization incident light leads to the transverse type, which is corresponding to poles of $r_{sp}$ (channel-III). In particular, the longitudinal SPPs have two different response channels: the channel-I SPP corresponds to the p-polarization reflection wave case, and the channel-II SPP corresponds to  the s-polarization reflection wave case. Besides, the transverse SPP is charaterized by p-polarization reflection wave. The channel-II and channel-III SPPs are caused by magneto-optical Kerr effect in TRS broken TIs surface states, while channel-I SPP is the ordinary SPP in common plasmonic materials.

 Considering the electrostatic limit condition $k_{0}<< k_{spp}$, where $k_{spp}$ is the plasmon wave vector, we deduce the dispersion for both two types of SPPs.

The dispersion of the longitudinal SPPs is

 \begin{equation}
 k^{L}_{spp}\simeq i(\epsilon+1)\omega/4\pi\sigma_{xx}
 \end{equation}
We can use the Drude formula of $\sigma_{xx}$ to reduce $k^{L}_{spp}$ as

\begin{equation}
k^{L}_{spp}\simeq (\hbar^{2}/e^{2}E_{F})(\epsilon+1)\omega (\omega+i/\tau)
\end{equation}
Here, $k^{L}_{spp}$ shows a quadratic dependence on the frequency $\omega$. The real part of $k^{L}_{spp}$ gives the wave length of longitudinal SPPs. The ratio of the wave lengths of the SPPs and free space light gives $\lambda^{L}_{spp}/\lambda_{0}\simeq [\alpha/(\epsilon+1)](E_{F}/\hbar\omega)$. The small ratio indicates the longitudinal SPPs exhibit tight confinement. Considering the damping mechanism of SPP due to  ohmic losses of electrons, the SPP have finite propagation distance $\delta^{L}_{spp}$, which is given by $1/2Im\{k^{L}_{spp}\}$. The characteristic lengths of longitudinal SPP stimulated by terahertz frequency photon are shown in Fig.3 (a)-(b).  The difference between the channel-I and channel-II SPPs can be observed when considering SPP-emitter coupling strength, as shown in section V.

Using the above method,  we obtain the dispersion of the transverse SPP:
\begin{equation}
 k^{T}_{spp} \simeq i(\epsilon-1)\omega/4\pi\sigma_{xx}
 \end{equation}
Inserting Drude formula of $\sigma_{xx}$ into Eq. (12) gives:
\begin{equation}
 k^{T}_{spp}\simeq (\hbar^{2}/e^{2}E_{F})(\epsilon-1)\omega (\omega+i/\tau)
\end{equation}
The ratio of the wave length of transverse SPP and free space light $\lambda^{T}_{spp}/\lambda_{0} \simeq  [\alpha/(\epsilon-1)](E_{F}/\hbar\omega)$
shows the degree of the SPP confinement. Similar to the longitudinal SPP case, the propagation distance $\delta^{T}_{spp}$ of transverse SPP is determined by $1/2Im\{ k^{T}_{spp}\}$. The corresponding characteristic lengths of transverse SPP are shown in Fig.3 (a)-(b). In this paper, we focus on the moderate frequency regime ($\hbar\omega\lesssim E_{F}$), thus we neglect the effect of retardation on transverse SPP. The transverse SPP is a characteristic of the massive helical surface states.

The SPPs can be experimentally measured by coupling polarized light to the surface of TIs. In order to excite a SPP, the frequency of the incident light must equal to the frequency of SPP. However, the SPP cannot be directly excited by electromagnetic field due to the conservation of parallel momentum in homogenous materials. The problem can be overcome by special experiment techniques such as confined geometries\cite{Fei2012,ChenJ2012} or artificial periodic structure.\cite{Feng2011,Pietro2013} The SPPs can be imaged by scanning near-field optical microscope in experiments.

\section{SPP-emitter coupling}

 We consider an experimental realizable setup with a quantum dipole as an emitter close to a gapped TI surface states. The emitter energy can be transfered into the plasmon collective channels along the TI surface. The reflected electric field outside the TI can be enhanced due to the SPPs. We use the decay rate $\Gamma$ to describe degree of the coupling strength between the dipole emitter
$\bm{\mu}$ and electric field, which involves the dipole field and the scattered field $\mathbf{E}_{s}$:\cite{Novotny}

\begin{equation}
\Gamma=\Gamma_{0}+\frac{2}{\hbar}Im\{\boldsymbol{\mu}^{*}\cdot \mathbf{E}_{s}\}
\end{equation}
where $\Gamma_{0}=4k_{0}^{3}|\bm{\mu}|^{2}/3\hbar$ is the contribution of the dipole electric field. The second term corresponds to the energy dissipation related to the scattered electric field $\mathbf{E}_{s}$. The decay rate $\Gamma$ can be expressed in terms of the Fresnel coefficients:\cite{Novotny}

\begin{equation}
\Gamma=\Gamma_{0}+\frac{1}{\hbar}\int^{\infty}_{0}k_{\parallel}dk_{\parallel}
 Re\{G_{s}\frac{e^{2ik_{\bot}z}}{k_{\bot}}\}
\end{equation}
where
\begin{equation}
G_{s}=|\mu_{\parallel}|^{2}k_{0}^{2}(r_{ss}+r_{ps})+(2|\mu_{\bot}|^{2}k_{\parallel}^{2}-|\mu_{\parallel}|^{2}k_{\bot}^{2})(r_{pp}+r_{sp})
\end{equation}
here $z$ is the distance between the dipole emitter and the TI surface, $\mu_{\parallel}$ and $\mu_{\bot}$ are the parallel and perpendicular components of the dipole moment $\bm{\mu}$.  In the electrostatic limit, the pole of different types of Fresnel coefficients dominates the decay rate.  Thus, the integral can be easily estimated by analyzing the contribution of the SPPs related to the pole of Fresnel coefficients. We calculate the decay rate of three SPP channels, respectively. For simplicity, we set $\hbar=c=1$.

\begin{figure}[htbp]
  \centering
  \includegraphics[width=8cm]{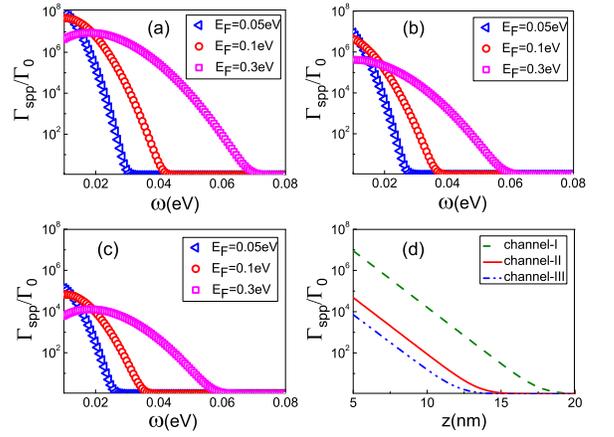}\\
  \caption{(color online) Decay rate of a dipole emitter coupling to TI SPPs. (a),(b) and (c) show contribution of channel-I, channel-II and channel-III SPPs as a function of photon energy for different Fermi energies $E_{F}$, respectively. In (a), (b) and (c) the dipole is placed 10 nm away from the TI surface. (d) gives out the distance dependence of emitter-SPP coupling. The Fermi energy $E_{F}$ is 0.1 eV, and the photon energy $\omega$ is 0.03 eV. The decay rate $\Gamma_{spp}$ is normalized to the the free-space value $\Gamma_{0}$  in all sub-figures. We set the gap energy $\triangle=20$ meV, the dipole orientation $\theta=45^{\circ}$, and dielectric constant $\varepsilon=100$ in all cases.}
\end{figure}

Although the channel-I and channel-II SPPs have the same dispersion, these two channels show different coupling strength with the emitter, as shown in Fig.4 (a)-(b). The channel-I SPP decay rate is:
\begin{equation}
\Gamma^{I}_{spp}=(|\mu_{\parallel}|^{2}+2|\mu_{\bot}|^{2})\frac{(2\pi)^{4}}{(\epsilon+1)}\frac{e^{-4\pi z/\lambda^{L}_{spp}}}{{\lambda^{L}_{spp}}^{3}}
\end{equation}
As shown in Fig.4 (a), the decay rate is peaked at certain photon energy $\omega_{c}$. From Eq. (11), the photon energy $\omega$ determines the SPP wave length $\lambda^{L}_{spp}$. When $z\ll\lambda^{L}_{spp}$, the decay rate is enhanced by $g=[3\pi f/2(\epsilon+1)](\lambda_{0}/\lambda^{L}_{spp})^{3}$ orders of magnitude, where
$f=3/2$ for dipole  orientation $\theta=45^{\circ}$. When $z>\lambda^{L}_{spp}$, the SPP decay rate $\Gamma^{I}_{spp}$ exponentially decreases. As shown in Fig.4 (a), the decay rate $\Gamma^{I}_{spp}$ decrease with enlarging $\omega$, due to the tight confinement ($\lambda^{L}_{spp}\ll\lambda_{0}$) of SPP for large $\omega$.

The decay rate of channel-II SPP is
\begin{equation}
\Gamma^{II}_{spp}=|\mu_{\parallel}|^{2}\frac{(2\pi)^{4}}{(\epsilon+1)^{2}}\frac{\triangle}{E_{F}\omega^{2}}\frac{e^{-4\pi z/\lambda^{L}_{sp}}}{{\lambda^{L}_{spp}}^{3}}
\end{equation}
The decay rate of channel-II SPP is shown in Fig.4(b). When $z\ll\lambda^{L}_{spp}$, the decay rate is enhanced by $g=[6\pi^{2} f \sigma_{xy}/(\epsilon+1)^{2}\omega k_{0}](\lambda_{0}/\lambda^{L}_{spp})^{3}$ orders of magnitude, where $f=1/2$ for dipole orientation $\theta=45^{\circ}$. When $z>\lambda_{sp}$, the SPP decay rate $\Gamma^{II}_{spp}$ decreases exponentially. Although the decay rate for channel-I and channel-II SPPs share common dependence on the photon energy, these two cases are quantitatively different. Comparing Fig.4 (a) and Fig.4 (b), we conclude the former channel is one order of magnitude larger than the latter one.

The decay rate of channel-III SPP is
\begin{equation}
\begin{aligned}
\Gamma^{III}_{spp}=&(|\mu_{\parallel}|^{2}+2|\mu_{\bot}|^{2})\frac{(2\pi)^{6}}{\epsilon-1}\frac{\triangle}{E_{F}}\frac{1}{8\pi^{2}/(\lambda_{spp}^{T})^{2}+(\epsilon-1)\omega^{2}}\\
&\times \frac{e^{-4\pi z/\lambda_{spp}^{T}}}{(\lambda_{spp}^{T})^{5}}
\end{aligned}
\end{equation}
The decay rate $\Gamma^{III}_{spp}$ is enhanced by $g=[\frac{8\pi^{4} f \sigma_{xy}}{8\pi^{2}/(\lambda_{spp}^{T})^{2}+(\epsilon-1)\omega k_{0}}](\lambda_{0})^{3}/(\lambda_{spp}^{T})^{5}$ orders of magnitude, where $f=3/2$ for dipole orientation $\theta=45^{\circ}$. When $z>\lambda_{spp}^{T}$, the SPP decay rate $\Gamma^{III}_{spp}$ exhibits an exponential damping, as shown in Fig. 4(c).

In all cases, the SPP decay rate is peaked at the order of $10^{4}\backsim 10^{7}$ for different channels in the terahertz frequency range. In addition, the decay rates rely on the distance between emitter and TI surface. As shown in Fig. 4(d), the decay rates decrease with enlarging the distance, characterized by a exponential behavior. Our simulation show that when a dipole emitter is placed close to TI surface, the decay rate is greatly enhanced and the emission energy is converted into SPPs in TI surface. This makes TI surface state a good platform for strong light-matter interactions.

\section{conclusion}

In conclusion, we consider plasmonic properties of gapped TI surface and investigate the SPPs on a TI-vacuum interface. When TRS is broken, the gapped TI surface states display magneto-optical Kerr effect. This mechanism give rise to both the transverse SPP and the longitudinal SPP. To be specific, there are three channels of SPPs, corresponding to poles of determinant of  Fresnel reflection matrix. The channel-I and channel-II SPP belong to longitudinal type, while the channel-III SPP is transverse type. Furthermore, we derive the dispersions of SPPs and find that all three SPP channels display tight confinement and relatively long propagation distances in terahertz frequency range. Finally, we consider the light-matter interactions between SPP channels and a dipole emitter. Our results show that TI can be a good platform for strong light-matter interactions.

\section*{ACKNOWLEDGMENTS}
We thank the insightful discussion with Jinhua Gao, Y. X. Xing and Hua Jiang. This work is financially supported by MOST of China (2012CB821402), NSF-China under Grants No. 91221302, and China Post-doctroal Science Foundation under Grant No. 2012M520099.

\end{document}